\documentclass[
 reprint,
 graphics,
 floatfix,
 tightenlines,
 superscriptaddress,
nofootinbib,
nobibnotes,
 amsmath,
 amssymb,
 aps,
pre
]{revtex4-1}

\usepackage{graphicx}
\usepackage{dcolumn}
\usepackage{bm}
\usepackage{amsmath}
\usepackage{subfig}
\DeclareMathOperator{\Tr}{tr}
\usepackage{color}
\usepackage[linktocpage,colorlinks=true,linkcolor=blue,citecolor=blue,breaklinks=true]{hyperref}
\usepackage{verbatim}
\usepackage{breakcites}
\newcommand{\ad}[1]{\textcolor{blue}{{\it#1}}}

\usepackage{ulem}

\renewcommand{\vec}[1]{\boldsymbol{#1}}

\begin{document}

\title{Electric-field induced shape transition of nematic tactoids}%

\author{Luuk Metselaar}
\affiliation{The Rudolf Peierls Centre for Theoretical Physics, 1 Keble Rd., Oxford OX1 3NP, UK}
\author{Ivan Dozov}
\affiliation{Laboratoire de Physique des Solides, Universit\'e Paris-Sud, Universit\'e Paris-Saclay, CNRS, UMR 8502, Orsay, France}
\author{Krassimira Antonova}
\affiliation{Institute of Solid State Physics, Bulgarian Academy of Sciences, Sofia, Bulgaria}
\author{Emmanuel Belamie}
\affiliation{Institut Charles Gerhardt Montpellier, ENSCM, Montpellier, France}
\author{Patrick Davidson}
\affiliation{Laboratoire de Physique des Solides, Universit\'e Paris-Sud, Universit\'e Paris-Saclay, CNRS, UMR 8502, Orsay, France}
\author{Julia M. Yeomans}
\author{Amin Doostmohammadi}
\affiliation{The Rudolf Peierls Centre for Theoretical Physics, 1 Keble Rd., Oxford OX1 3NP, UK}

\date{\today}

\begin{abstract}

The occurrence of new textures of liquid crystals is an important factor in tuning their optical and photonics properties. Here, we show, both experimentally and by numerical computation, that under an electric field chitin tactoids (i.e.~nematic droplets) can stretch to aspect ratios of more than $15$, leading to a transition from a spindle-like to a cigar-like shape. We argue that the large extensions occur because the elastic contribution to the free energy is dominated by the anchoring. We demonstrate that the elongation involves hydrodynamic flow and is reversible, the tactoids return to their original shapes upon removing the field.

\end{abstract}

\pacs{Valid PACS appear here}
\maketitle

\section{Introduction}
When dispersed in a solvent, a wide variety of chemical and natural substances, from polypeptides to viral particles and intracellular actin filaments, spontaneously organise in liquid-crystalline nematic phases. The nematic phase usually appears within the disordered isotropic phase through the nucleation and growth of birefringent droplets, termed {\em tactoids} \cite{Zocher1925}.

\begin{figure}[b]
\centering
\includegraphics[width=\linewidth]{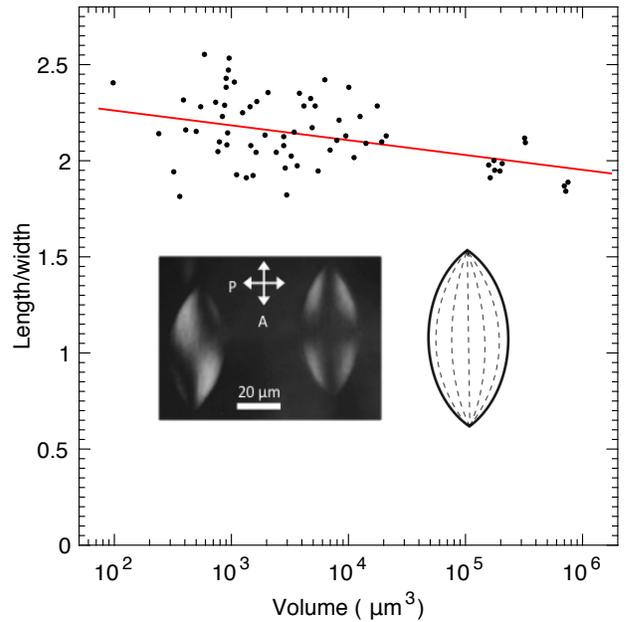}
\caption{Aspect ratio of the tactoids as a function of the tactoid volume before applying the electric field. Left inset: the observation under crossed polarisers (white double headed arrows) reveals the bipolar structure of the tactoids. The two tactoids look different because the axis of the tactoid on the right is exactly parallel to the analyser direction whereas that of the tactoid on the left lies at an angle of $\sim$ 3 $^{\circ}$. Right inset: schematic drawing of a tactoid, with dashed lines representing the director field.}
\label{fig:aspect_ratio}
\end{figure}

These nematic droplets have characteristic shapes that have already been the subject of several theoretical studies \cite{Kaznacheev2002,Prinsen2003,Lettinga2005,Lettinga2006,VanBijnen2012,Otten2012,Everts2016} and experiments on systems such as vanadium pentoxide (V$_2$O$_5$) \cite{Sonin1998,Kaznacheev2002}, carbon nanotubes \cite{Puech2010,Jamali2015}, rod-like viruses \cite{Dogic2003,Modlinska2015}, F-actin in cells \cite{Oakes2007}, chromonic liquid crystals \cite{Kim2013}, and cellulose nano-crystals \cite{Revol1994,Park2014,Wang2016}. These investigations showed that the typical tactoid shapes, for example the so-called spindle-like shape with a bipolar director field (see the right inset in Fig.~\ref{fig:aspect_ratio}), are determined by the balance between a low surface tension, strong anchoring, and the elasticity of these lyotropic nematics. Such distinctive shapes are not usually found in the case of thermotropic nematics that have very large surface tension, leading to almost spherical drops, even though the director field can be bipolar \cite{Volovik1983}. 

Like most liquid crystals, tactoids are very sensitive to external magnetic or electric fields but only a few studies have so far dealt with this phenomenon. The most obvious effect of the field on a tactoid is usually the overall orientation of its long axis with respect to the field direction \cite{Kaznacheev2002}. Subjecting tactoids of plate-like colloidal particles to sufficiently large magnetic fields induces a sudden rearrangement of the director field and leads to stable split-core defect structures \cite{Verhoeff2011}, while tactoids of rod-like colloids can also be slightly stretched without qualitative change of the internal structure of the director field \cite{Kaznacheev2002}.

Here we present experimental results to show that, for large electric fields, chitin tactoids with a bipolar director field can elongate to aspect ratios of more than $15$, forming cigar-shaped droplets with a uniform director configuration except for at the tips. We find that the elongation process is entirely reversible upon removal of the electric field to recreate the original tactoid shapes. By comparing the experiments and numerical simulations we argue that the substantial elongations are the results of weak elasticity and strong anchoring.

\section{Materials and Methods}
Chitin extracted and purified from crab shells was kindly provided by Katakura Chikkarin Co. Ltd. as a white flaky powder. Chitin flakes (1 g) were hydrolysed in a round flask with 20 mL 3 M HCl at the boil for 90 minutes and then quickly cooled down on ice. After thorough dialysis against deionised water until a pH of about 4 was reached, dispersion of the particles was achieved by sonication (using a Branson sonicator) for 30 minutes. Remaining aggregates were eliminated by low speed centrifugation. The suspensions were then concentrated by dialysis against a 50 gL$^{-1}$ solution of PEG 35000 gmol$^{-1}$ in water acidified with HCl (pH 4). Stock suspensions were brought to a concentration of about 10\% in weight, measured precisely by evaporation and weighing. Final dilution was achieved by adding to the stock suspensions the appropriate amount (in weight) of the corresponding aqueous solvent. The chitin suspension had a concentration of circa 3 wt\%, which corresponds to the biphasic isotropic-nematic region of the phase diagram, close to the isotropic border \cite{Belamie2004}. Considering a density of 1.425 gcm$^{-3}$ for chitin nanocrystals, the chitin volume fraction in the suspension is thus $\phi$ $\approx$ 0.02. This value represents the overall volume fraction of the suspension which actually demixes into coexisting isotropic and nematic phases, with respective volume fractions $\phi_\text{I}$ = 0.018 and $\phi_\text{N}$ = 0.028.

For electro-optical studies, samples were filled into optically flat glass capillaries (VitroCom) of thickness 200 $\mu$m that were flame-sealed. Nematic tactoids form and slowly sediment over a few hours as the macroscopic isotropic-nematic phase separation proceeds. The response of the tactoids to the electric field was measured during sedimentation to avoid their interaction with the inner surfaces of the capillary.

The set-up used for the application of the electric field in the sample has been described in detail elsewhere \cite{Dozov2011}. A high frequency (typically f = 300-700 kHz) sinusoidal a.c.~voltage with amplitude U$_0$ was applied to the suspension using two external electrodes made up from aluminium foil and placed directly on the outside wall of the capillary with inter-electrode gap L$_e$ = 1.06 mm. This geometry of the experiment enabled us to apply strong and uniform electric fields without direct contact of the electrodes with the suspension, avoiding electrolysis and other electrochemical artifacts and minimising the Joule heating of the sample. The r.m.s. value of the field penetrating in the sample is E$_\text{in}$ = C$_\text{d}$C$_\text{s}$(f)E$_\text{rms}$, where E$_\text{rms}$ = U$_0$/($\sqrt{2}$L$_\text{e}$) is the externally applied field, and the coefficients C$_\text{d}$ $\approx$ 0.48 and 0 $<$ C$_\text{s}$(f) $<$ 1 take into account respectively the field attenuation due to the dielectric mismatch between the capillary glass wall and the suspension, and the screening of the field by the conductivity charges in the suspension. During the experiment, the sample was placed on the stage of a polarising microscope Olympus BX 51 and the evolution of the tactoids was observed and periodically photographed using a CCD camera for the measurement of their length L and diameter D as a function of the time and field strength. When possible, the evolution of each tactoid was followed individually from one photograph to the next one. Otherwise, for long times and/or strong fields, the follow-up was only statistical, because, in these conditions, the individual tracking was hindered by the convective flow due to the Joule heating of the sample.

Two distinct series of experiments were performed on similar batches of chitin aqueous suspensions. Qualitatively the results are similar, but in the first series the response to the field was stronger and saturated at lower frequency than in the second series. The frequency and the strength of the applied field were therefore set to slightly different values in the two experiments in order to maximise the response of the sample to the field and to minimise the Joule heating and the related convective flows. In the first series, the frequency of the applied voltage was f = 300 kHz (with C$_\text{s}$ $\approx$ 1) and the maximum field strength was E$_\text{rms}$ = 160 Vmm$^{-1}$; in the second series, the frequency was f = 700 kHz (with C$_\text{s}$ $\approx$ 0.7) and the maximum field strength was E$_\text{rms}$ = 450 Vmm$^{-1}$. The optimal frequency was determined in an auxiliary study of the field-induced nematic order in the isotropic phase of the same colloidal dispersion. The results were not influenced by the frequency variation when the frequency-dependent losses for penetration of the field in the sample were taken into account.

The frequency dependence of the screening coefficient C$_\text{s}$(f) was measured in the isotropic phase of the suspensions. From the same experiment, the nematic order parameter \cite{Dozov2011} induced by the maximum field of 450 Vmm$^{-1}$ was estimated to be only S = 0.04, which confirms that the main effect of the electric field is to align the nematic director inside the tactoid, with only a minor perturbation of the isotropic fluid around it.

\begin{figure*}[t]
\centering
\includegraphics[width=0.75\linewidth]{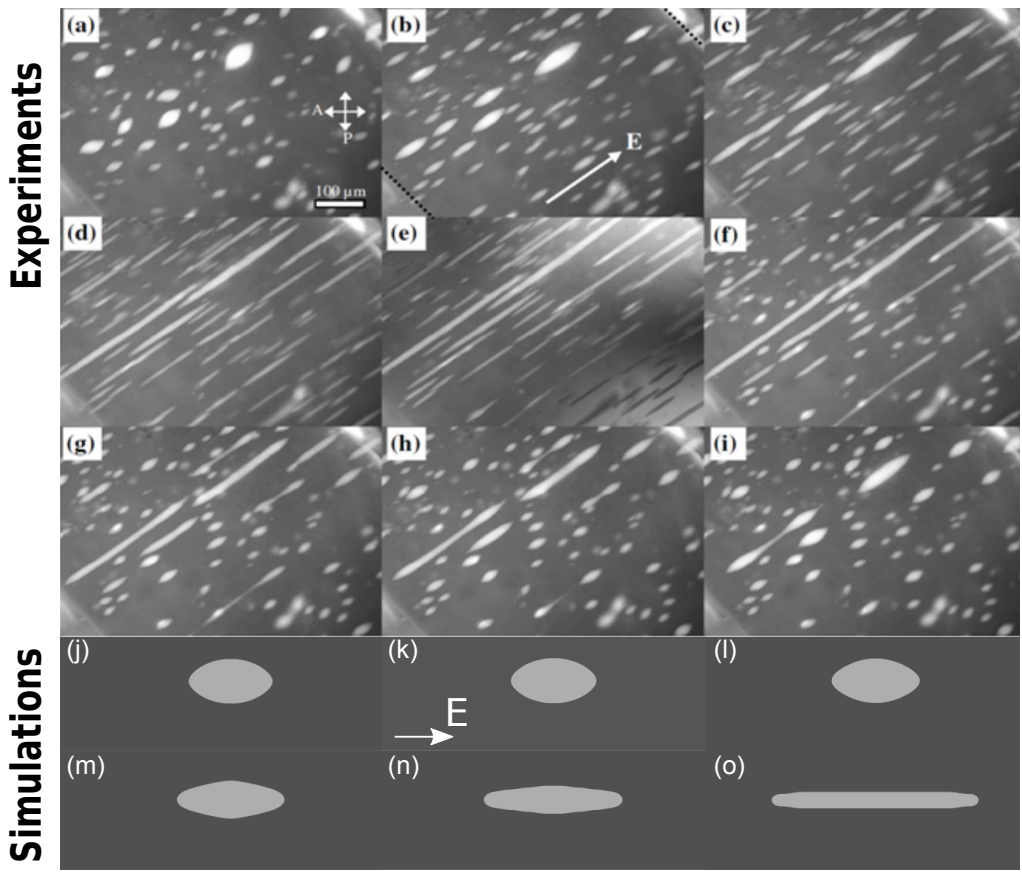}
\caption{Time-evolution of the tactoids under a field E$_{\text{rms}}$ =160 Vmm$^{-1}$, f=300 kHz, and relaxation after switching off the field. Experiments: (a--e) The field is applied respectively for time $\tau_{\text{on}}$= 0, 120, 210, 640, 1080 seconds. (f--i) The sample has relaxed for $\tau_{\text{off}}$ = 230, 380, 530, 820 seconds after the field removal. The double-headed arrows in (a) show the orientations of the polariser P and the analyser A. In (b) the field direction $\vec{E}$ (white arrow) and the shadows cast by the out-of-focus electrodes (black dotted lines) are shown. Note that in (e) the contrast varies and is inverted in some regions, due to introducing a Berek compensator. Simulation: (j--o) Time evolution of simulated tactoids under an electric field. (j) Equilibrium tactoid shape. The director field is bipolar and the aspect ratio is $\sim$2. Because we employ a diffuse interface model the tips may appear slightly rounded. (k--o) The tactoid 200, 400, 800, 1600 and 30000 simulation time steps after applying the electric field. The droplet stretches to an aspect ratio of $\sim$16, and becomes cylindrical with conical tips.}
\label{fig:experimental_snapshots}
\end{figure*}
\begin{figure}[b]
\centering
\includegraphics[width=1.0\linewidth]{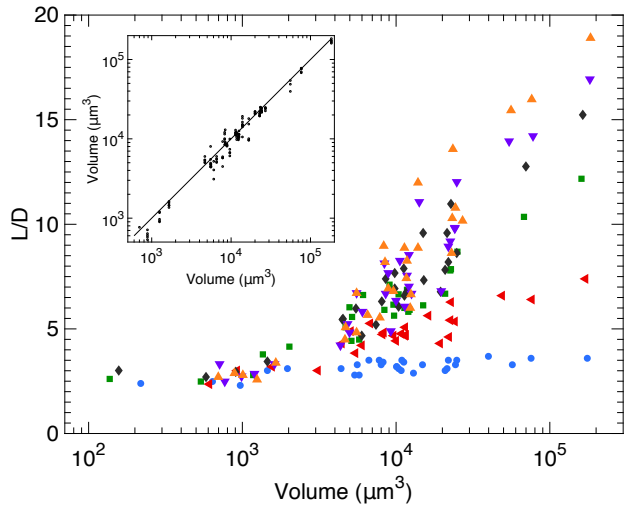}
\caption{Axial ratio of the tactoids as a function of their volume and the time under field E$_\text{rms}$ = 160 Vmm$^{-1}$ and f = 300 kHz. The data corresponding to the same time under field are plotted with the same symbol. Blue circles are taken after 120 s, red triangles after 210 s, green squares after 410 s, black diamonds after 640 s, purple inverted triangles after 850 s and orange triangles after 1080 s. The data measured for the same tactoid lie almost on a vertical line, as the volume of the tactoid is approximately independent of the time under the field. This is demonstrated by the inset which shows the volume of the tactoids at different times as a function of the final volume after 1080 seconds under field.}\label{fig:experiment_volume}
\end{figure}

\begin{figure}[t]
\centering
\includegraphics[width=1.0\linewidth]{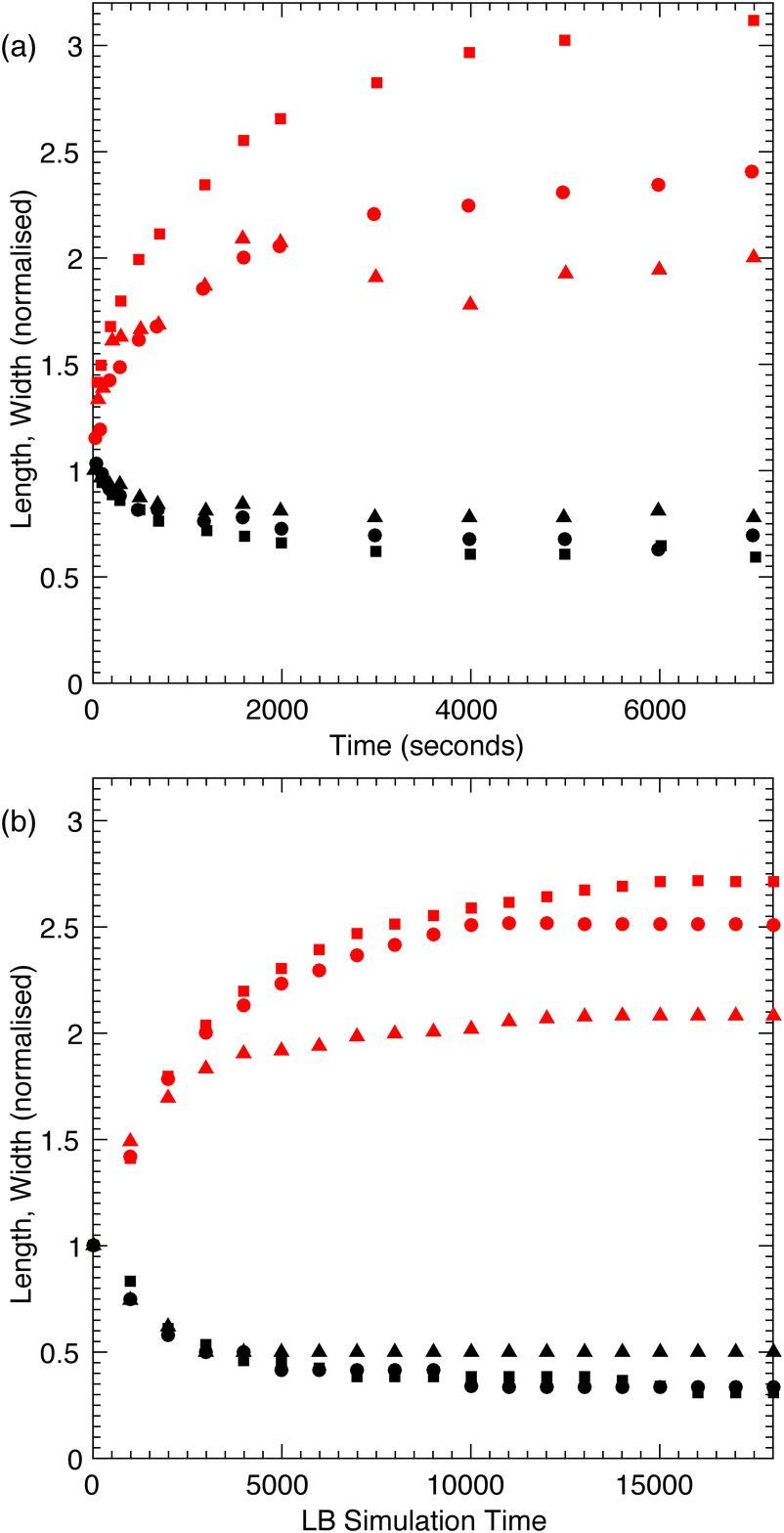}
\caption{Time-dependence of length L (red markers) and width D (black markers) of tactoids of different sizes. L and D are normalised by their initial values in the absence of an electric field. (a) Experiments: Three chitin tactoids with LD$^2 = 2.0 \times 10^4$ $\mu$m$^3$ (triangles), $1.5 \times 10^5$ $\mu$m$^3$ (dots) and $3.7 \times 10^5$ $\mu$m$^3$ (squares), subject to a 500 kHz a.c. field E$_{\text{rms}}$ = 154 Vmm$^{-1}$. (b) Simulations: Three tactoids with LD = $\pi$8$^2$  (triangles), $\pi$14$^2$ (dots) and $\pi$16$^2$ (squares).}\label{fig:stretch}
\end{figure}

\section{Experimental Results}
In the absence of any applied field, the tactoids that spontaneously form upon the isotropic-nematic transition have a spindle-like shape and bipolar morphology, independent of their size (see the insets in Fig.~\ref{fig:aspect_ratio}). They are randomly oriented throughout the sample. Moreover, their aspect ratio, L/D $\approx$ 2 where L is the length and D the diameter, is only weakly dependent on the volume $V$ (see Fig.~\ref{fig:aspect_ratio}). The interfacial anchoring and the surface tension scale with the surface area of the tactoids, while the elasticity scales with the volume. The constant aspect ratio observed here therefore suggests that the interfacial anchoring and the surface tension dominate in this system. We note that this is in contrast to previous works, where aspect ratios of tactoids were shown to decrease with increasing volume \cite{Bernal1941,Kaznacheev2002,Modlinska2015,Jamali2015}, in reasonable agreement with theoretical predictions \cite{Kaznacheev2002,Kaznacheev2003,Prinsen2003,Jamali2015} for models where elasticity is relevant. 

Before the first application of the field, the tactoids are completely disoriented. When the electric field is applied, a reorientation of the director along the field is observed inside the tactoid, except in the close vicinity of the nematic-isotropic interface. This purely elastic process is very fast (response times $<$1s) and is reversible upon field removal. At much longer time scales (10-100 seconds) the tactoids slowly reorient along the field, rotating approximately as rigid bodies. Simultaneously, but at even longer time scales (100-1000 seconds), the tactoids become distorted as their aspect ratio increases. Therefore, short pulses of electric field, separated by relaxation times, were first applied to orient all the tactoids along the electric field direction and avoid any distortion of their shape. A population of tactoids of very different volumes but of same shape and orientation was thus produced (Fig.~\ref{fig:experimental_snapshots}a). 

When submitted to an a.c.~electric field of constant amplitude, the tactoids strongly deformed and became highly elongated along the field direction (Fig.~\ref{fig:experimental_snapshots}b-e). With time, the largest tactoids evolved from the bipolar structure to a tubular shape with conical tips (Fig.~\ref{fig:experimental_snapshots}d-e). In the simulations, described in more detail below, this behaviour is nicely recovered (see Fig.~\ref{fig:experimental_snapshots}j-o). The deformation is reversible as the tactoids became bipolar again and their aspect ratio returned to its initial value in $\sim$10$^3$s after the electric field was switched off (Fig.~\ref{fig:experimental_snapshots}f-i). This behaviour is significantly different from previous work on tactoids where the aspect ratio only increased by $\sim$10\% under a 1T magnetic field and the tactoid structure remained strictly bipolar \cite{Sonin1998,Kaznacheev2002,Kaznacheev2003}.

The deformation of the tactoids in the field strongly depends on their volume (see Fig.~\ref{fig:experiment_volume}). For example, small tactoids of volume 10$^2$ - 10$^3$ $\mu$m$^3$ did not deform much on the time scale of the experiment (30 minutes), while the aspect ratio of large tactoids (10$^5$ $\mu$m$^3$) doubled in 5 minutes and reached L/D $\approx$ 15 in 30 minutes. The volume of the tactoids remained almost constant during the deformation process (see the inset in Fig.~\ref{fig:experiment_volume}).  

The evolution of the tactoid shapes when a field is applied is summarised in Fig.~\ref{fig:stretch}a which shows the time evolution of the tactoid length and diameter for different tactoid volumes. Small tactoids undergo small deformations but on the relatively fast time scale of a few hundred seconds. By contrast, the largest tactoids show very large deformations occurring on time scales of several thousands of seconds. The large time scale suggests that flows are involved in the deformation of large tactoids under the electric field. The same qualitative behaviour was observed for field amplitudes between 100 and 450 Vmm$^{-1}$.

\begin{figure}[t]
\centering
\includegraphics[width=1.0\linewidth]{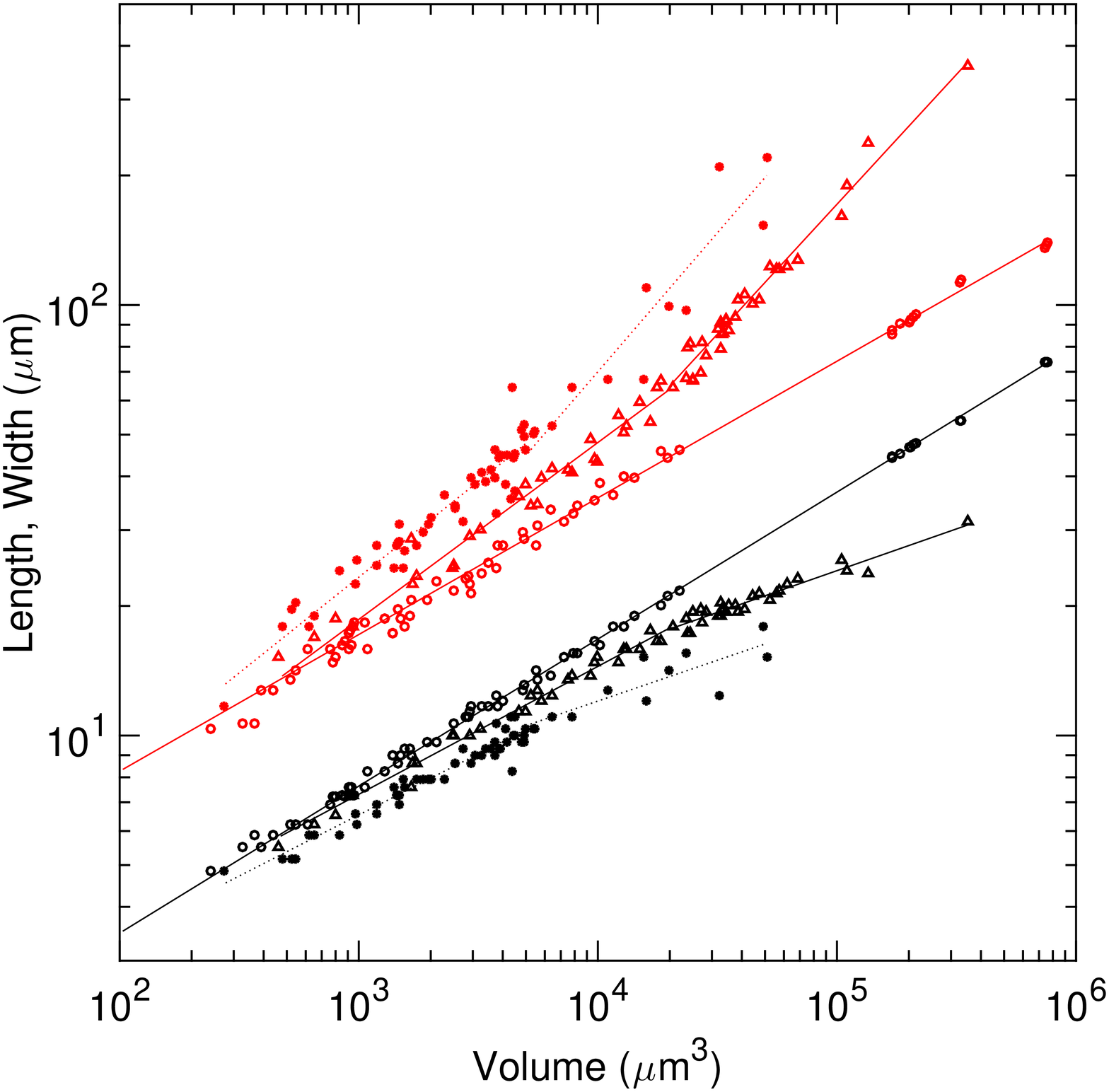}
\caption{Length L (red markers) and width D (black markers) of the tactoids as a function of the tactoid volume LD$^2$ before the application of the field (open circles), after 61000 s under E$_\text{rms}$ = 154 Vmm$^{-1}$ (triangles) and after 3600 s under E$_\text{rms}$ = 309 Vmm$^{-1}$ (closed symbols). The no-field data are fitted with a power law with coefficients $\beta_\text{L}$ = 0.32 and $\beta_\text{D}$ = 0.34, very close to the V$^{1/3}$ dependence expected for constant aspect ratio L/D. The 154 Vmm$^{-1}$ data are piecewise fitted with power laws with respectively $\beta_\text{L}$ = 0.41 and 0.60, and $\beta_\text{D}$ = 0.30 and 0.20, with the crossover at LD$^2$ $\sim$ 2x10$^4$ $\mu$m$^3$. For E$_\text{rms}$ = 309 Vmm$^{-1}$, the coefficients are respectively $\beta_\text{L}$ 0.44 and 0.64, and $\beta_\text{D}$ = 0.28 and 0.19, with the crossover at LD$^2$ $\sim$ 5x10$^3$ $\mu$m$^3$.}\label{fig:scaling}
\end{figure}

Without a field, the length and diameter of the tactoids scale with the cube root of their volume (see Fig.~\ref{fig:scaling}), which is expected because all tactoids have the same aspect ratio. Under the electric field, the tactoid diameter decreases with increasing tactoid volume, showing a crossover from a cube root behaviour to a weaker dependence as a function of volume. The value of the crossover volume, V$_\text{c}$, depends on the electric field amplitude; it is V$_\text{c}$ $\sim $2x10$^4$ $\mu$m$^3$ at E$_\text{rms}$ = 154 Vmm$^{-1}$ and V$_\text{c}$ $\sim$ 5x10$^3$ $\mu$m$^3$ at E$_\text{rms}$ = 309 Vmm$^{-1}$. There is a larger corresponding increase in the tactoid length leading to aspect ratios as high as 15. Moreover, we observed the qualitative change of the tactoid morphology from the usual spindle-like shape with bipolar structure to a tubular shape. This consists of an almost cylindrical central region with uniform director distribution ending with two conical tips with an approximately bipolar director field.

To our knowledge, these features have not been reported in any other system. For comparison, under a 1T magnetic field, large V$_2$O$_5$ tactoids are aligned along the field and are slightly elongated, with an aspect ratio increasing by only about 10\%, without any change of their bipolar structure \cite{Kaznacheev2002}. In a similar way, the axial ratio of lens-like tactoids formed by gibbsite platelets only slightly varies (by $\sim$30\%) under a 1T magnetic field \cite{Verhoeff2011}, in good agreement with theoretical models \cite{Verhoeff2011,Otten2012}.

The action of the magnetic field is due to the coupling of the field with the magnetic anisotropy of the nematic phase; this effect can only be expected if (at least) one of the coexisting phases is anisotropic. This is no longer the case when an electric field is applied to biphasic systems. Apart from the direct coupling of the electric field with the dielectric anisotropy of the nematic phase, given by the energy term $\epsilon_0\Delta\epsilon$(\textbf{E.n})$^2$/2, where $\Delta\epsilon$ = $\epsilon_\parallel$ - $\epsilon_\bot$ is the dielectric anisotropy, there are additional energy terms that are unrelated to the phase anisotropy. These terms are relevant even for coexisting isotropic phases. They are due to the different dielectric constant ($\epsilon_\text{in}$ $\neq$ $\epsilon_\text{out}$) and conductivity (K$_\text{in}$ $\neq$ K$_\text{out}$) inside and outside of the droplet interface. Under a field, electric charges accumulate on the interface, respectively as bound (polarisation) or mobile (conductivity) charges. Their surface density is anisotropic and depends on the orientation and shape of the droplet \cite{Okonski1953,Okonski1957,Melcher1969,Vizika1992,Saville1993}. Therefore, an initially spherical isotropic droplet elongated under field, becomes ellipsoidal, and orients with its main axis along the field. This process minimises the electric energy cost at the price of increasing the interface area and the interface energy. This phenomenon is well known for isotropic droplets and, when the dielectric and/or conductivity contrast is large, a droplet elongation of 15\% has been reported \cite{Vizika1992}.

The same mechanisms are involved in the electric-field-induced elongation of thermotropic nematic droplets in coexistence with their isotropic melt or with another liquid \cite{Park1994,Lev2000,Lev2001,Auernhammer2009}. Usually for thermotropic nematics, the surface tension is much larger than the anchoring energy and the droplets are (approximately) spherical without a field. Under a field, the nematic director in the droplet is aligned along the field (assuming that $\epsilon_\parallel$ $>$ $\epsilon_\bot$) and the effective dielectric and conductivity contrasts at the interface are now respectively $\epsilon_\parallel$-$\epsilon_\text{out}$ and K$_\parallel$-K$_\text{out}$. When these contrasts are strong and the surface tension of the nematic is relatively weak (as usual), the droplet aspect ratio can be quite large, up to 3, especially for low frequencies, f $<$ 1 kHz, where the conductivity mechanism is dominant \cite{Park1994,Lev2001}. However, even in this case, the shape of the droplet is not spindle-like but ellipsoidal because, as in the isotropic case, there is only competition between the electric and surface tension energies, the anchoring energy being negligible. Yet, the flow induced by the friction drag exerted by the electric current in the liquid is an artifact that can induce cusp-like distortions of the droplet in d.c. or very low frequency a.c. fields \cite{Lev2001}.

All these auxiliary mechanisms are irrelevant in our experiment. Indeed, working with external electrodes, without direct contact with the liquids, requires the use of very high frequencies to achieve enough penetration of the field in the liquid. We use frequencies much higher than the relaxation frequency of the conductivity (Maxwell-Wagner mechanism \cite{Wagner1914}) so that this effect can be neglected. For the same reason, the flow directly induced by the electric current drag is also negligible. Moreover, the dielectric contrast $\delta\epsilon$ $\approx$ $\epsilon_\text{in}$-$\epsilon_\text{out}$ is very small ($\delta\epsilon$ $<<$ $\epsilon_\text{out}$) in chitin suspensions and its influence on the tactoid shape is negligibly small.

The only remaining electric-field contribution to the tactoid energy is the direct coupling with the dielectric anisotropy of the nematic phase, like the magnetic field case. Nevertheless, for large tactoids, deformations were observed two orders of magnitude larger than reported for V$_2$O$_5$ in a magnetic field.

The most unexpected result is the field-induced transition from a spindle-like to a tubular shape for large tactoids. Theoretical models explain the spindle-like shape by the competition between the bend and splay elastic energies, the surface tension and the anchoring energy \cite{Kaznacheev2002,Prinsen2003,Kaznacheev2003}. Under application of an external field the energy balance is modified, leading to variation of the tactoid aspect ratio. However, the models predict that the shape should remain spindle-like, as in the absence of field, contrary to our observations.

At least at a qualitative level, one does not expect significant differences in elastic properties and surface tension between chitin suspensions and other aqueous lyotropic colloids of rod-like particles. However, we argued above that the independence of the tactoid shape on its volume suggests that the elastic contribution to the free energy is small compared to the surface tension and anchoring contributions. Moreover close examination of the orientation of the director in the bulk and at the interface of the tubular tactoids in strong fields revealed that although the director orientation is completely uniform in the bulk (and parallel to the electric field), close to the tips the director remains strictly parallel to the surface and relaxes over a length of about 3 $\mu$m. This demonstrates that the anchoring is actually very strong and any tilt of the director from the surface can be neglected at the droplet boundary, even in strong fields.

Since, at this stage, the physical origin of this novel shape remains unknown and since its description seems out of reach of the analytical models available to date, numerical simulations were employed. Moreover, the complex dynamical evolution of the tactoids in an electric field is difficult to model analytically but it can be captured by a numerical approach.

\section{Numerical Simulations}
In order to understand the nature of the shape change of the tactoids and to scrutinise the underpinning physical mechanism for their large elongation under electric fields, we employ a nematohydrodynamics approach. We use a hybrid lattice Boltzmann-finite difference method \cite{DoostmohammadiPRL2016} to solve the equations of motion describing a mixture of a nematic liquid crystal with an isotropic fluid, and work in two dimensions.
The symmetric traceless tensor $\textbf{Q}$ is used to describe the nematic order of the fluid \cite{DeGennes1995}. The second order parameter used in the model is the liquid crystal concentration $\varphi$. The isotropic and nematic phases coexist, and the amount of each is conserved, as observed experimentally. The free energy of the system is
\begin{equation}
\mathcal{F} = \int \left\{f \big(\varphi,\nabla \varphi,\textbf{Q},\nabla \textbf{Q}\big) - \mu\varphi \right\}d^2\textbf{r}
\end{equation}
where
\begin{equation}
\begin{split}
f = \frac{1}{2}A_{\varphi}\varphi^2\left(1-\varphi^2\right) + A_0\varphi^2\Big\{\frac{1}{2}\Big(1-\frac{\eta(\varphi)}{3}\Big)\Tr(\bm{Q}^2) \\ 
- \frac{\eta(\varphi)}{3}\Tr(\bm{Q}^3) + \frac{\eta(\varphi)}{4}\Tr(\bm{Q}^2)^2\Big\} +\frac{1}{2}K\left(\bm{\nabla Q}\right)^2 \\
- \epsilon_a\bm{E}\cdot \bm{Q} \cdot \bm{E} + \frac{1}{2}k_\varphi\left(\bm{\nabla}\varphi\right)^2 + L_0 \bm{\nabla}\varphi \cdot\bm{Q}\cdot \bm{\nabla}\varphi,
\end{split}
\end{equation}
where $A_\varphi$, $A_0$, $K$, $\epsilon_a$ and $k_\varphi$ are all positive constants. The first term in $f$ is the bulk free energy for a binary fluid, with equilibria at $\varphi = 0,1$ corresponding to the isotropic and nematic phases, respectively \cite{Orlandini1995a}. The second term describes a first-order, isotropic-nematic transition at $\eta(\varphi)=2.7$ \cite{Matsuyama2002}. This contribution is weighted by $\varphi^2$, since only the nematic phase contributes in this way to the free energy. The third term is the Frank elastic free energy density. We use the one elastic constant approximation, corresponding to having the splay and bend Frank elastic constants equal. This approximation holds well since, as discussed earlier, and confirmed numerically, the weak dependence of the  aspect ratio on the tactoid volume (Fig.~\ref{fig:aspect_ratio}) indicates that the tactoid formation is primarily controlled by the competition of the anchoring on the surface of the nematic droplet and the interfacial tension between the nematic and isotropic phases.
The fourth term is the contribution from an electric field $\textbf{E}$, with $\epsilon_\text{a}$ the dielectric anisotropy. The fifth term penalises gradients in the concentration and provides an interfacial tension. The last term couples the concentration gradient to the liquid crystal orientation, modelling interfacial anchoring of the liquid crystal. $L_0 > 0$ encourages the director to align parallel to the isotropic--nematic interface (planar anchoring), whereas $L_0 < 0$ encourages perpendicular (homeotropic) anchoring \cite{Sulaiman2006}. To reproduce the experimental results qualitatively, we choose $A_0$=1.5, $A_\varphi$=0.28, $K$=0.093, $\epsilon_\text{a}$=1, $k_\varphi$=0.05 and $L_0$=0.06. 

The evolution of the order parameters $\varphi$ and $\textbf{Q}$ is dictated by the Cahn-Hilliard and Beris-Edwards equations respectively \cite{Cahn1958},\cite{Beris1994},
\begin{equation}\label{eq:CahnHilliard}
\partial_t \varphi + \bm{\nabla} \cdot \left(\varphi\bm{u}\right) = M \bm{\nabla}^2 \mu,
\end{equation}
\begin{equation}\label{eq:BerisEdwards}
\left(\partial_t + \bm{u} \cdot \bm{\nabla}\right)\bm{Q} - \bm{S} = \Gamma \bm{H}\ad{,}
\end{equation}
where $\bm{S} = \left(\xi \bm{D}+\bm{\Omega}\right)\left(\bm{Q}+\frac{1}{3}\bm{I}\right) + \left(\bm{Q}+\frac{1}{3}\bm{I}\right)\left(\xi \bm{D}-\bm{\Omega}\right)-2\xi\left(\bm{Q}+\frac{1}{3}\bm{I}\right)\Tr\left(\bm{Q}\bm{W}\right)$ characterises the response of the nematic tensor to velocity gradients \cite{Beris1994}. $\bm{D}$ and $\bm{\Omega}$ are the symmetric and antisymmetric parts of the velocity gradient tensor $W_{\alpha\beta}=\partial_\beta u_\alpha$, and $\xi$ is the tumbling parameter. $\Gamma$ is a rotational diffusivity and $\bm{H} = -\left(\frac{\delta\mathcal{F}}{\delta\bm{Q}} - \frac{1}{3}\bm{I}\Tr\frac{\delta\mathcal{F}}{\delta\bm{Q}}\right)$ is the molecular field, modelling the relaxation towards a free energy minimum while $\bm{Q}$ remains traceless. In equation (\ref{eq:CahnHilliard}) $M$ is the mobility and $\mu = \frac{\partial f}{\partial \varphi} - \bm{\nabla} \cdot\left(\frac{\partial f}{\partial (\bm{\nabla} \varphi)}\right)$ is the chemical potential.

The velocity $\bm{u}$ evolves according to the generalised Navier-Stokes equations
\begin{align}\label{eq:NavierStokes}
\nabla\cdot\bm{u}&=0,\\
\rho\left(\partial_t + \bm{u} \cdot \bm{\nabla}\right)\bm{u} &= \bm{\nabla} \cdot \bm{\Pi},
\end{align}
where $\bm{\Pi}$ is the stress tensor. The stress consists of a viscous stress $\bm{\Pi}_{viscous}=2\eta\bm{D}$, an elastic stress $\bm{\Pi}_{elastic}=-P_0\bm{I}-\xi\bm{H}\left(\bm{Q}+\frac{1}{3}\bm{I}\right)-\xi\left(\bm{Q}+\frac{1}{3}\bm{I}\right)\bm{H}+2\xi\left(\bm{Q}+\frac{1}{3}\bm{I}\right)\Tr\left(\bm{QH}\right)+\bm{Q}\bm{H}-\bm{H}\bm{Q}- \bm{\nabla Q}\delta f/\delta\bm{\nabla Q}$, and a capillary stress $\bm{\Pi}_{cap}=\left(f-\mu\varphi\right)\bm{I}-\bm{\nabla}\varphi\left(\delta\mathcal{F}/\delta\bm{\nabla}\varphi\right)$, where $\eta$ is the fluid viscosity and $P_0$ is the bulk pressure. We use $M$=0.5, $\Gamma$=0.7, $\xi$=0.7 and $P_0$=0.25 to obtain qualitative agreement with the experiments. 

The simulations are set up starting with a circular nematic droplet, with a uniform director field and with radius between 8 and 16 lattice units, situated on a 240 $\times$ 120 rectangular lattice. Before the electric field is turned on the droplet is equilibrated for 64000 time steps. It relaxes to a shape with sharp tips, an aspect ratio of about two, and a bipolar director field, recovering the characteristic properties of the experimentally observed tactoids (Fig.~\ref{fig:experimental_snapshots}j).  When the field is turned on, the tactoid becomes highly extended, taking on a cylindrical shape with conical tips as shown by the series of snapshots in Fig.~\ref{fig:experimental_snapshots}k-o. This shape is in agreement with the experiments (Fig.~\ref{fig:experimental_snapshots}e).

Simulation results for the time evolution of tactoids of different area are compared to experiment in Fig.~\ref{fig:stretch}, showing similar behaviour. When hydrodynamic flow was not included in the simulations it was not possible to regain the dynamics observed in the experiments, providing evidence that the long time-scales observed in the stretching of the larger tactoids are of hydrodynamic origin. 

The experimental observation that the tactoid shape is independent of its volume suggests that the elastic contribution to the free energy balance is small compared to those from the surface tension and anchoring in the chitin system. For simulation parameters which reproduce the experimental results the free energy contributions due to anchoring and surface tension are similar, whereas that from the elasticity is about ten times lower in the zero field configuration.

\begin{figure}[t]
\centering
\includegraphics[width=\linewidth]{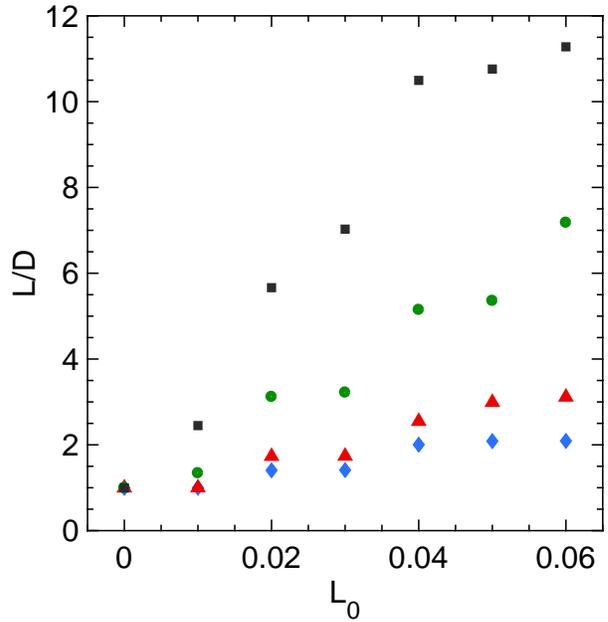}
\caption{Simulation results showing how the final aspect ratio L/D of a tactoid depends on its anchoring strength L$_0$ and the electric field strength. Blue diamonds represent an electric field strength of 0.06, red triangles 0.10, green circles 0.15 and black squares 0.20.}
\label{fig:anchoring}
\end{figure}

Having validated the simulations against the experiments we can use them to investigate how the final tactoid extension depends on the anchoring strength and the strength of the applied electric field. The results, shown in Fig.~\ref{fig:anchoring}, show that the extension increases strongly with anchoring strength. In zero field tactoids with zero anchoring strength are round with a uniform director field and they are unperturbed by the field, as expected. For larger anchoring strengths (0.04 to 0.06 in simulation units) the tactoids become spindle-like with a bipolar director configuration in zero field. Applying an electric field leads to increases in the aspect ratio, of up to a few hundred percent for $L_0=0.06$.

We saw no evidence of elongation at field strengths below 0.04 (in simulation units), independent of the tactoid volume, suggesting that there is a small energy barrier to the initiation of stretching. To check that this was not due to the lattice pinning the long, straight interface we repeated the calculations for spherical droplets and found the same qualitative behaviour. We hypothesise that the energy barrier occurs because the strongly anchored directors at the interface need a minimum field to be able to distort the interface so that they can align with the field direction. This is supported by the fact that the energy barrier decreases with decreasing anchoring.

\section{Summary}
In summary, we have shown experimentally and by numerical simulations that the application of an electric field to nematic tactoids induces a transition from the common spindle-like shape with a bipolar director field to a cigar-like shape with an almost uniform director field. The drops extend in the direction of the field to reach aspect ratios of over ten. We argue that this is a consequence of the low elasticity and strong anchoring of the chitin tactoids.

From a practical point of view, subjecting tactoids to shear flow has been shown to improve the optical properties of liquid crystal films by enhancing the alignment of the particles \cite{Park2014}, and using electric fields can achieve this same effect in a much more controllable environment. Indeed, when doped with acrylamide monomers, these biphasic aqueous chitin suspensions could easily be polymerised under electric field, as was recently demonstrated in the case of clay suspensions \cite{Paineau2012}.

To further investigate the morphological properties of liquid crystals, and their tunability, it would be interesting to look at the coalescence of tactoids. In the experiments it was occasionally observed that tactoids may coalesce while elongating, to break up again during retraction. Since bigger tactoids react more strongly to an electric field, coalescence may be an important factor in improving alignment. Furthermore chitin tactoids offer an easily accessible confined liquid crystal environment for studying fundamental properties of liquid crystal solutions. \\
~\\
This project has received funding from the European Union's Horizon 2020 research and innovation programme under the DiStruc Marie Sk\l{}odowska-Curie grant agreement No. 641839, the ERC Advanced Grant MiCE 291234, and the Agence Nationale pour la Recherche ANR (France) grant NASTAROD. We thank Paul van der Schoot and Pavlik Lettinga for helpful discussions.
\bibliographystyle{apsrev4-1}
\bibliography{library}

\begin{thebibliography}{42}%
\makeatletter
\providecommand \@ifxundefined [1]{%
 \@ifx{#1\undefined}
}%
\providecommand \@ifnum [1]{%
 \ifnum #1\expandafter \@firstoftwo
 \else \expandafter \@secondoftwo
 \fi
}%
\providecommand \@ifx [1]{%
 \ifx #1\expandafter \@firstoftwo
 \else \expandafter \@secondoftwo
 \fi
}%
\providecommand \natexlab [1]{#1}%
\providecommand \enquote  [1]{``#1''}%
\providecommand \bibnamefont  [1]{#1}%
\providecommand \bibfnamefont [1]{#1}%
\providecommand \citenamefont [1]{#1}%
\providecommand \href@noop [0]{\@secondoftwo}%
\providecommand \href [0]{\begingroup \@sanitize@url \@href}%
\providecommand \@href[1]{\@@startlink{#1}\@@href}%
\providecommand \@@href[1]{\endgroup#1\@@endlink}%
\providecommand \@sanitize@url [0]{\catcode `\\12\catcode `\$12\catcode
  `\&12\catcode `\#12\catcode `\^12\catcode `\_12\catcode `\%12\relax}%
\providecommand \@@startlink[1]{}%
\providecommand \@@endlink[0]{}%
\providecommand \url  [0]{\begingroup\@sanitize@url \@url }%
\providecommand \@url [1]{\endgroup\@href {#1}{\urlprefix }}%
\providecommand \urlprefix  [0]{URL }%
\providecommand \Eprint [0]{\href }%
\providecommand \doibase [0]{http://dx.doi.org/}%
\providecommand \selectlanguage [0]{\@gobble}%
\providecommand \bibinfo  [0]{\@secondoftwo}%
\providecommand \bibfield  [0]{\@secondoftwo}%
\providecommand \translation [1]{[#1]}%
\providecommand \BibitemOpen [0]{}%
\providecommand \bibitemStop [0]{}%
\providecommand \bibitemNoStop [0]{.\EOS\space}%
\providecommand \EOS [0]{\spacefactor3000\relax}%
\providecommand \BibitemShut  [1]{\csname bibitem#1\endcsname}%
\let\auto@bib@innerbib\@empty
\bibitem [{\citenamefont {Zocher}(1925)}]{Zocher1925}%
  \BibitemOpen
  \bibfield  {author} {\bibinfo {author} {\bibfnamefont {H.}~\bibnamefont
  {Zocher}},\ }\href@noop {} {\bibfield  {journal} {\bibinfo  {journal}
  {Zeitschrift f{\"u}r Anorganische und Allgemeine Chemie}\ }\textbf {\bibinfo
  {volume} {147}},\ \bibinfo {pages} {91} (\bibinfo {year} {1925})}\BibitemShut
  {NoStop}%
\bibitem [{\citenamefont {Kaznacheev}\ \emph {et~al.}(2002)\citenamefont
  {Kaznacheev}, \citenamefont {Bogdanov},\ and\ \citenamefont
  {Taraskin}}]{Kaznacheev2002}%
  \BibitemOpen
  \bibfield  {author} {\bibinfo {author} {\bibfnamefont {A.~V.}\ \bibnamefont
  {Kaznacheev}}, \bibinfo {author} {\bibfnamefont {M.~M.}\ \bibnamefont
  {Bogdanov}}, \ and\ \bibinfo {author} {\bibfnamefont {S.~A.}\ \bibnamefont
  {Taraskin}},\ }\href@noop {} {\bibfield  {journal} {\bibinfo  {journal} {J.
  Exp. Theor. Phys.}\ }\textbf {\bibinfo {volume} {95}},\ \bibinfo {pages} {57}
  (\bibinfo {year} {2002})}\BibitemShut {NoStop}%
\bibitem [{\citenamefont {Prinsen}\ and\ \citenamefont {{van der
  Schoot}}(2003)}]{Prinsen2003}%
  \BibitemOpen
  \bibfield  {author} {\bibinfo {author} {\bibfnamefont {P.}~\bibnamefont
  {Prinsen}}\ and\ \bibinfo {author} {\bibfnamefont {P.}~\bibnamefont {{van der
  Schoot}}},\ }\href@noop {} {\bibfield  {journal} {\bibinfo  {journal} {Phys.
  Rev. E}\ }\textbf {\bibinfo {volume} {68}},\ \bibinfo {pages} {021701}
  (\bibinfo {year} {2003})}\BibitemShut {NoStop}%
\bibitem [{\citenamefont {Lettinga}\ \emph {et~al.}(2005)\citenamefont
  {Lettinga}, \citenamefont {Kang}, \citenamefont {Imhof}, \citenamefont
  {Derks},\ and\ \citenamefont {Dhont}}]{Lettinga2005}%
  \BibitemOpen
  \bibfield  {author} {\bibinfo {author} {\bibfnamefont {M.~P.}\ \bibnamefont
  {Lettinga}}, \bibinfo {author} {\bibfnamefont {K.}~\bibnamefont {Kang}},
  \bibinfo {author} {\bibfnamefont {A.}~\bibnamefont {Imhof}}, \bibinfo
  {author} {\bibfnamefont {D.}~\bibnamefont {Derks}}, \ and\ \bibinfo {author}
  {\bibfnamefont {J.~K.~G.}\ \bibnamefont {Dhont}},\ }\href@noop {} {\bibfield
  {journal} {\bibinfo  {journal} {J. Phys. Condens. Matter}\ }\textbf {\bibinfo
  {volume} {17}},\ \bibinfo {pages} {S3609} (\bibinfo {year}
  {2005})}\BibitemShut {NoStop}%
\bibitem [{\citenamefont {Lettinga}\ \emph {et~al.}(2006)\citenamefont
  {Lettinga}, \citenamefont {Kang}, \citenamefont {Holmqvist}, \citenamefont
  {Imhof}, \citenamefont {Derks},\ and\ \citenamefont {Dhont}}]{Lettinga2006}%
  \BibitemOpen
  \bibfield  {author} {\bibinfo {author} {\bibfnamefont {M.~P.}\ \bibnamefont
  {Lettinga}}, \bibinfo {author} {\bibfnamefont {K.}~\bibnamefont {Kang}},
  \bibinfo {author} {\bibfnamefont {P.}~\bibnamefont {Holmqvist}}, \bibinfo
  {author} {\bibfnamefont {A.}~\bibnamefont {Imhof}}, \bibinfo {author}
  {\bibfnamefont {D.}~\bibnamefont {Derks}}, \ and\ \bibinfo {author}
  {\bibfnamefont {J.~K.~G.}\ \bibnamefont {Dhont}},\ }\href@noop {} {\bibfield
  {journal} {\bibinfo  {journal} {Phys. Rev. E}\ }\textbf {\bibinfo {volume}
  {73}},\ \bibinfo {pages} {011412} (\bibinfo {year} {2006})}\BibitemShut
  {NoStop}%
\bibitem [{\citenamefont {{van Bijnen}}\ \emph {et~al.}(2012)\citenamefont
  {{van Bijnen}}, \citenamefont {Otten},\ and\ \citenamefont {{van der
  Schoot}}}]{VanBijnen2012}%
  \BibitemOpen
  \bibfield  {author} {\bibinfo {author} {\bibfnamefont {R.~M.~W.}\
  \bibnamefont {{van Bijnen}}}, \bibinfo {author} {\bibfnamefont {R.~H.~J.}\
  \bibnamefont {Otten}}, \ and\ \bibinfo {author} {\bibfnamefont
  {P.}~\bibnamefont {{van der Schoot}}},\ }\href@noop {} {\bibfield  {journal}
  {\bibinfo  {journal} {Phys. Rev. E}\ }\textbf {\bibinfo {volume} {86}},\
  \bibinfo {pages} {051703} (\bibinfo {year} {2012})}\BibitemShut {NoStop}%
\bibitem [{\citenamefont {Otten}\ and\ \citenamefont {{van der
  Schoot}}(2012)}]{Otten2012}%
  \BibitemOpen
  \bibfield  {author} {\bibinfo {author} {\bibfnamefont {R.~H.~J.}\
  \bibnamefont {Otten}}\ and\ \bibinfo {author} {\bibfnamefont
  {P.}~\bibnamefont {{van der Schoot}}},\ }\href@noop {} {\bibfield  {journal}
  {\bibinfo  {journal} {J. Chem. Phys.}\ }\textbf {\bibinfo {volume} {137}},\
  \bibinfo {pages} {154901} (\bibinfo {year} {2012})}\BibitemShut {NoStop}%
\bibitem [{\citenamefont {Everts}\ \emph {et~al.}(2016)\citenamefont {Everts},
  \citenamefont {Punter}, \citenamefont {Samin}, \citenamefont {{van der
  Schoot}},\ and\ \citenamefont {{van Roij}}}]{Everts2016}%
  \BibitemOpen
  \bibfield  {author} {\bibinfo {author} {\bibfnamefont {J.}~\bibnamefont
  {Everts}}, \bibinfo {author} {\bibfnamefont {M.}~\bibnamefont {Punter}},
  \bibinfo {author} {\bibfnamefont {S.}~\bibnamefont {Samin}}, \bibinfo
  {author} {\bibfnamefont {P.}~\bibnamefont {{van der Schoot}}}, \ and\
  \bibinfo {author} {\bibfnamefont {R.}~\bibnamefont {{van Roij}}},\
  }\href@noop {} {\bibfield  {journal} {\bibinfo  {journal} {J. Chem. Phys.}\
  }\textbf {\bibinfo {volume} {144}},\ \bibinfo {pages} {194901} (\bibinfo
  {year} {2016})}\BibitemShut {NoStop}%
\bibitem [{\citenamefont {Sonin}(1998)}]{Sonin1998}%
  \BibitemOpen
  \bibfield  {author} {\bibinfo {author} {\bibfnamefont {A.~S.}\ \bibnamefont
  {Sonin}},\ }\href@noop {} {\bibfield  {journal} {\bibinfo  {journal} {J.
  Mater. Chem.}\ }\textbf {\bibinfo {volume} {8}},\ \bibinfo {pages} {2557}
  (\bibinfo {year} {1998})}\BibitemShut {NoStop}%
\bibitem [{\citenamefont {Puech}\ \emph {et~al.}(2010)\citenamefont {Puech},
  \citenamefont {Grelet}, \citenamefont {Poulin}, \citenamefont {Blanc},\ and\
  \citenamefont {{van der Schoot}}}]{Puech2010}%
  \BibitemOpen
  \bibfield  {author} {\bibinfo {author} {\bibfnamefont {N.}~\bibnamefont
  {Puech}}, \bibinfo {author} {\bibfnamefont {E.}~\bibnamefont {Grelet}},
  \bibinfo {author} {\bibfnamefont {P.}~\bibnamefont {Poulin}}, \bibinfo
  {author} {\bibfnamefont {C.}~\bibnamefont {Blanc}}, \ and\ \bibinfo {author}
  {\bibfnamefont {P.}~\bibnamefont {{van der Schoot}}},\ }\href@noop {}
  {\bibfield  {journal} {\bibinfo  {journal} {Phys. Rev. E}\ }\textbf {\bibinfo
  {volume} {82}},\ \bibinfo {pages} {020702} (\bibinfo {year}
  {2010})}\BibitemShut {NoStop}%
\bibitem [{\citenamefont {Jamali}\ \emph {et~al.}(2015)\citenamefont {Jamali},
  \citenamefont {Behabtu}, \citenamefont {Senyuk}, \citenamefont {Lee},
  \citenamefont {Smalyukh}, \citenamefont {{van der Schoot}},\ and\
  \citenamefont {Pasquali}}]{Jamali2015}%
  \BibitemOpen
  \bibfield  {author} {\bibinfo {author} {\bibfnamefont {V.}~\bibnamefont
  {Jamali}}, \bibinfo {author} {\bibfnamefont {N.}~\bibnamefont {Behabtu}},
  \bibinfo {author} {\bibfnamefont {B.}~\bibnamefont {Senyuk}}, \bibinfo
  {author} {\bibfnamefont {J.~A.}\ \bibnamefont {Lee}}, \bibinfo {author}
  {\bibfnamefont {I.~I.}\ \bibnamefont {Smalyukh}}, \bibinfo {author}
  {\bibfnamefont {P.}~\bibnamefont {{van der Schoot}}}, \ and\ \bibinfo
  {author} {\bibfnamefont {M.}~\bibnamefont {Pasquali}},\ }\href@noop {}
  {\bibfield  {journal} {\bibinfo  {journal} {Phys. Rev. E}\ }\textbf {\bibinfo
  {volume} {91}},\ \bibinfo {pages} {042507} (\bibinfo {year}
  {2015})}\BibitemShut {NoStop}%
\bibitem [{\citenamefont {Dogic}(2003)}]{Dogic2003}%
  \BibitemOpen
  \bibfield  {author} {\bibinfo {author} {\bibfnamefont {Z.}~\bibnamefont
  {Dogic}},\ }\href@noop {} {\bibfield  {journal} {\bibinfo  {journal} {Phys.
  Rev. Lett.}\ }\textbf {\bibinfo {volume} {91}},\ \bibinfo {pages} {165701}
  (\bibinfo {year} {2003})}\BibitemShut {NoStop}%
\bibitem [{\citenamefont {Modlinska}\ \emph {et~al.}(2015)\citenamefont
  {Modlinska}, \citenamefont {Alsayed},\ and\ \citenamefont
  {Gibaud}}]{Modlinska2015}%
  \BibitemOpen
  \bibfield  {author} {\bibinfo {author} {\bibfnamefont {A.}~\bibnamefont
  {Modlinska}}, \bibinfo {author} {\bibfnamefont {A.~M.}\ \bibnamefont
  {Alsayed}}, \ and\ \bibinfo {author} {\bibfnamefont {T.}~\bibnamefont
  {Gibaud}},\ }\href@noop {} {\bibfield  {journal} {\bibinfo  {journal} {Nat.
  Sci. Rec.}\ }\textbf {\bibinfo {volume} {5}},\ \bibinfo {pages} {18}
  (\bibinfo {year} {2015})}\BibitemShut {NoStop}%
\bibitem [{\citenamefont {Oakes}\ \emph {et~al.}(2007)\citenamefont {Oakes},
  \citenamefont {Viamontes},\ and\ \citenamefont {Tang}}]{Oakes2007}%
  \BibitemOpen
  \bibfield  {author} {\bibinfo {author} {\bibfnamefont {P.~W.}\ \bibnamefont
  {Oakes}}, \bibinfo {author} {\bibfnamefont {J.}~\bibnamefont {Viamontes}}, \
  and\ \bibinfo {author} {\bibfnamefont {J.~X.}\ \bibnamefont {Tang}},\
  }\href@noop {} {\bibfield  {journal} {\bibinfo  {journal} {Phys. Rev. E}\
  }\textbf {\bibinfo {volume} {75}},\ \bibinfo {pages} {061902} (\bibinfo
  {year} {2007})}\BibitemShut {NoStop}%
\bibitem [{\citenamefont {Kim}\ \emph {et~al.}(2013)\citenamefont {Kim},
  \citenamefont {Shiyanovskii},\ and\ \citenamefont {Lavrentovich}}]{Kim2013}%
  \BibitemOpen
  \bibfield  {author} {\bibinfo {author} {\bibfnamefont {Y.-K.}\ \bibnamefont
  {Kim}}, \bibinfo {author} {\bibfnamefont {S.~V.}\ \bibnamefont
  {Shiyanovskii}}, \ and\ \bibinfo {author} {\bibfnamefont {O.~D.}\
  \bibnamefont {Lavrentovich}},\ }\href@noop {} {\bibfield  {journal} {\bibinfo
   {journal} {J. Phys. Condens. Matter}\ }\textbf {\bibinfo {volume} {25}},\
  \bibinfo {pages} {404202} (\bibinfo {year} {2013})}\BibitemShut {NoStop}%
\bibitem [{\citenamefont {Revol}\ \emph {et~al.}(1994)\citenamefont {Revol},
  \citenamefont {Godbout}, \citenamefont {Dong}, \citenamefont {Gray},
  \citenamefont {Chanzy},\ and\ \citenamefont {Maret}}]{Revol1994}%
  \BibitemOpen
  \bibfield  {author} {\bibinfo {author} {\bibfnamefont {J.-F.}\ \bibnamefont
  {Revol}}, \bibinfo {author} {\bibfnamefont {L.}~\bibnamefont {Godbout}},
  \bibinfo {author} {\bibfnamefont {X.-M.}\ \bibnamefont {Dong}}, \bibinfo
  {author} {\bibfnamefont {D.~G.}\ \bibnamefont {Gray}}, \bibinfo {author}
  {\bibfnamefont {H.}~\bibnamefont {Chanzy}}, \ and\ \bibinfo {author}
  {\bibfnamefont {G.}~\bibnamefont {Maret}},\ }\href@noop {} {\bibfield
  {journal} {\bibinfo  {journal} {Liq. Cryst.}\ }\textbf {\bibinfo {volume}
  {16}},\ \bibinfo {pages} {1} (\bibinfo {year} {1994})}\BibitemShut {NoStop}%
\bibitem [{\citenamefont {Park}\ \emph {et~al.}(2014)\citenamefont {Park},
  \citenamefont {Noh}, \citenamefont {Schutz}, \citenamefont {Salazar-Alvarez},
  \citenamefont {Scalia}, \citenamefont {Bergstrom},\ and\ \citenamefont
  {Lagerwall}}]{Park2014}%
  \BibitemOpen
  \bibfield  {author} {\bibinfo {author} {\bibfnamefont {J.~H.}\ \bibnamefont
  {Park}}, \bibinfo {author} {\bibfnamefont {J.}~\bibnamefont {Noh}}, \bibinfo
  {author} {\bibfnamefont {C.}~\bibnamefont {Schutz}}, \bibinfo {author}
  {\bibfnamefont {G.}~\bibnamefont {Salazar-Alvarez}}, \bibinfo {author}
  {\bibfnamefont {G.}~\bibnamefont {Scalia}}, \bibinfo {author} {\bibfnamefont
  {L.}~\bibnamefont {Bergstrom}}, \ and\ \bibinfo {author} {\bibfnamefont
  {J.~P.~F.}\ \bibnamefont {Lagerwall}},\ }\href@noop {} {\bibfield  {journal}
  {\bibinfo  {journal} {Chem. Phys. Chem.}\ }\textbf {\bibinfo {volume} {15}},\
  \bibinfo {pages} {1477} (\bibinfo {year} {2014})}\BibitemShut {NoStop}%
\bibitem [{\citenamefont {Wang}\ \emph {et~al.}(2016)\citenamefont {Wang},
  \citenamefont {Hamad},\ and\ \citenamefont {MacLachlan}}]{Wang2016}%
  \BibitemOpen
  \bibfield  {author} {\bibinfo {author} {\bibfnamefont {P.-X.}\ \bibnamefont
  {Wang}}, \bibinfo {author} {\bibfnamefont {W.~Y.}\ \bibnamefont {Hamad}}, \
  and\ \bibinfo {author} {\bibfnamefont {M.~J.}\ \bibnamefont {MacLachlan}},\
  }\href@noop {} {\bibfield  {journal} {\bibinfo  {journal} {Nat. Commun.}\
  }\textbf {\bibinfo {volume} {7}},\ \bibinfo {pages} {11515} (\bibinfo {year}
  {2016})}\BibitemShut {NoStop}%
\bibitem [{\citenamefont {Volovik}\ and\ \citenamefont
  {Lavrentovich}(1983)}]{Volovik1983}%
  \BibitemOpen
  \bibfield  {author} {\bibinfo {author} {\bibfnamefont {G.~E.}\ \bibnamefont
  {Volovik}}\ and\ \bibinfo {author} {\bibfnamefont {O.~D.}\ \bibnamefont
  {Lavrentovich}},\ }\href@noop {} {\bibfield  {journal} {\bibinfo  {journal}
  {Zh. Eksp. Teor. Fiz.}\ }\textbf {\bibinfo {volume} {85}},\ \bibinfo {pages}
  {6} (\bibinfo {year} {1983})}\BibitemShut {NoStop}%
\bibitem [{\citenamefont {Verhoeff}\ \emph {et~al.}(2011)\citenamefont
  {Verhoeff}, \citenamefont {Otten}, \citenamefont {{van der Schoot}},\ and\
  \citenamefont {Lekkerkerker}}]{Verhoeff2011}%
  \BibitemOpen
  \bibfield  {author} {\bibinfo {author} {\bibfnamefont {A.~A.}\ \bibnamefont
  {Verhoeff}}, \bibinfo {author} {\bibfnamefont {R.~H.~J.}\ \bibnamefont
  {Otten}}, \bibinfo {author} {\bibfnamefont {P.}~\bibnamefont {{van der
  Schoot}}}, \ and\ \bibinfo {author} {\bibfnamefont {H.~N.~W.}\ \bibnamefont
  {Lekkerkerker}},\ }\href@noop {} {\bibfield  {journal} {\bibinfo  {journal}
  {J. Chem. Phys.}\ }\textbf {\bibinfo {volume} {134}},\ \bibinfo {pages}
  {044904} (\bibinfo {year} {2011})}\BibitemShut {NoStop}%
\bibitem [{\citenamefont {Belamie}\ \emph {et~al.}(2004)\citenamefont
  {Belamie}, \citenamefont {Davidson},\ and\ \citenamefont
  {Giraud-Guille}}]{Belamie2004}%
  \BibitemOpen
  \bibfield  {author} {\bibinfo {author} {\bibfnamefont {E.}~\bibnamefont
  {Belamie}}, \bibinfo {author} {\bibfnamefont {P.}~\bibnamefont {Davidson}}, \
  and\ \bibinfo {author} {\bibfnamefont {M.~M.}\ \bibnamefont
  {Giraud-Guille}},\ }\href@noop {} {\bibfield  {journal} {\bibinfo  {journal}
  {J. Phys. Chem. B}\ }\textbf {\bibinfo {volume} {108}},\ \bibinfo {pages}
  {39} (\bibinfo {year} {2004})}\BibitemShut {NoStop}%
\bibitem [{\citenamefont {Dozov}\ \emph {et~al.}(2011)\citenamefont {Dozov},
  \citenamefont {Paineau}, \citenamefont {Davidson}, \citenamefont {Antonova},
  \citenamefont {Baravian}, \citenamefont {Bihannic},\ and\ \citenamefont
  {Michot}}]{Dozov2011}%
  \BibitemOpen
  \bibfield  {author} {\bibinfo {author} {\bibfnamefont {I.}~\bibnamefont
  {Dozov}}, \bibinfo {author} {\bibfnamefont {E.}~\bibnamefont {Paineau}},
  \bibinfo {author} {\bibfnamefont {P.}~\bibnamefont {Davidson}}, \bibinfo
  {author} {\bibfnamefont {K.}~\bibnamefont {Antonova}}, \bibinfo {author}
  {\bibfnamefont {C.}~\bibnamefont {Baravian}}, \bibinfo {author}
  {\bibfnamefont {I.}~\bibnamefont {Bihannic}}, \ and\ \bibinfo {author}
  {\bibfnamefont {L.~J.}\ \bibnamefont {Michot}},\ }\href@noop {} {\bibfield
  {journal} {\bibinfo  {journal} {J. Phys. Chem. B}\ }\textbf {\bibinfo
  {volume} {115}},\ \bibinfo {pages} {24} (\bibinfo {year} {2011})}\BibitemShut
  {NoStop}%
\bibitem [{\citenamefont {Bernal}\ and\ \citenamefont
  {Fankuchen}(1941)}]{Bernal1941}%
  \BibitemOpen
  \bibfield  {author} {\bibinfo {author} {\bibfnamefont {J.}~\bibnamefont
  {Bernal}}\ and\ \bibinfo {author} {\bibfnamefont {I.}~\bibnamefont
  {Fankuchen}},\ }\href@noop {} {\bibfield  {journal} {\bibinfo  {journal} {J.
  Gen. Physiol.}\ }\textbf {\bibinfo {volume} {25}},\ \bibinfo {pages} {1}
  (\bibinfo {year} {1941})}\BibitemShut {NoStop}%
\bibitem [{\citenamefont {Kaznacheev}\ \emph {et~al.}(2003)\citenamefont
  {Kaznacheev}, \citenamefont {Bogdanov},\ and\ \citenamefont
  {Sonin}}]{Kaznacheev2003}%
  \BibitemOpen
  \bibfield  {author} {\bibinfo {author} {\bibfnamefont {A.}~\bibnamefont
  {Kaznacheev}}, \bibinfo {author} {\bibfnamefont {M.}~\bibnamefont
  {Bogdanov}}, \ and\ \bibinfo {author} {\bibfnamefont {A.}~\bibnamefont
  {Sonin}},\ }\href@noop {} {\bibfield  {journal} {\bibinfo  {journal} {J. Exp.
  Theor. Phys.}\ }\textbf {\bibinfo {volume} {97}},\ \bibinfo {pages} {1159}
  (\bibinfo {year} {2003})}\BibitemShut {NoStop}%
\bibitem [{\citenamefont {O'Konski}\ and\ \citenamefont
  {Thacher~Jr}(1953)}]{Okonski1953}%
  \BibitemOpen
  \bibfield  {author} {\bibinfo {author} {\bibfnamefont {C.~T.}\ \bibnamefont
  {O'Konski}}\ and\ \bibinfo {author} {\bibfnamefont {H.~C.}\ \bibnamefont
  {Thacher~Jr}},\ }\href@noop {} {\bibfield  {journal} {\bibinfo  {journal} {J.
  Phys. Chem.}\ }\textbf {\bibinfo {volume} {57}},\ \bibinfo {pages} {9}
  (\bibinfo {year} {1953})}\BibitemShut {NoStop}%
\bibitem [{\citenamefont {O'Konski}\ and\ \citenamefont
  {Harris}(1957)}]{Okonski1957}%
  \BibitemOpen
  \bibfield  {author} {\bibinfo {author} {\bibfnamefont {C.~T.}\ \bibnamefont
  {O'Konski}}\ and\ \bibinfo {author} {\bibfnamefont {F.~E.}\ \bibnamefont
  {Harris}},\ }\href@noop {} {\bibfield  {journal} {\bibinfo  {journal} {J.
  Phys. Chem.}\ }\textbf {\bibinfo {volume} {61}},\ \bibinfo {pages} {9}
  (\bibinfo {year} {1957})}\BibitemShut {NoStop}%
\bibitem [{\citenamefont {Melcher}\ and\ \citenamefont
  {Taylor}(1969)}]{Melcher1969}%
  \BibitemOpen
  \bibfield  {author} {\bibinfo {author} {\bibfnamefont {J.}~\bibnamefont
  {Melcher}}\ and\ \bibinfo {author} {\bibfnamefont {G.}~\bibnamefont
  {Taylor}},\ }\href@noop {} {\bibfield  {journal} {\bibinfo  {journal} {Annu.
  Rev. Fluid Mech.}\ }\textbf {\bibinfo {volume} {1}},\ \bibinfo {pages} {1}
  (\bibinfo {year} {1969})}\BibitemShut {NoStop}%
\bibitem [{\citenamefont {Vizika}\ and\ \citenamefont
  {Saville}(1992)}]{Vizika1992}%
  \BibitemOpen
  \bibfield  {author} {\bibinfo {author} {\bibfnamefont {O.}~\bibnamefont
  {Vizika}}\ and\ \bibinfo {author} {\bibfnamefont {D.}~\bibnamefont
  {Saville}},\ }\href@noop {} {\bibfield  {journal} {\bibinfo  {journal} {J.
  Fluid Mech.}\ }\textbf {\bibinfo {volume} {239}},\ \bibinfo {pages} {1}
  (\bibinfo {year} {1992})}\BibitemShut {NoStop}%
\bibitem [{\citenamefont {Saville}(1993)}]{Saville1993}%
  \BibitemOpen
  \bibfield  {author} {\bibinfo {author} {\bibfnamefont {D.}~\bibnamefont
  {Saville}},\ }\href@noop {} {\bibfield  {journal} {\bibinfo  {journal} {Phys.
  Rev. Lett.}\ }\textbf {\bibinfo {volume} {71}},\ \bibinfo {pages} {2907}
  (\bibinfo {year} {1993})}\BibitemShut {NoStop}%
\bibitem [{\citenamefont {Park}\ \emph {et~al.}(1994)\citenamefont {Park},
  \citenamefont {Clark},\ and\ \citenamefont {Noble}}]{Park1994}%
  \BibitemOpen
  \bibfield  {author} {\bibinfo {author} {\bibfnamefont {C.~S.}\ \bibnamefont
  {Park}}, \bibinfo {author} {\bibfnamefont {N.~A.}\ \bibnamefont {Clark}}, \
  and\ \bibinfo {author} {\bibfnamefont {R.~D.}\ \bibnamefont {Noble}},\
  }\href@noop {} {\bibfield  {journal} {\bibinfo  {journal} {Phys. Rev. Lett.}\
  }\textbf {\bibinfo {volume} {72}},\ \bibinfo {pages} {1838} (\bibinfo {year}
  {1994})}\BibitemShut {NoStop}%
\bibitem [{\citenamefont {Lev}\ \emph {et~al.}(2000)\citenamefont {Lev},
  \citenamefont {Nazarenko}, \citenamefont {Nych},\ and\ \citenamefont
  {Tomchuk}}]{Lev2000}%
  \BibitemOpen
  \bibfield  {author} {\bibinfo {author} {\bibfnamefont {B.}~\bibnamefont
  {Lev}}, \bibinfo {author} {\bibfnamefont {V.}~\bibnamefont {Nazarenko}},
  \bibinfo {author} {\bibfnamefont {A.}~\bibnamefont {Nych}}, \ and\ \bibinfo
  {author} {\bibfnamefont {P.}~\bibnamefont {Tomchuk}},\ }\href@noop {}
  {\bibfield  {journal} {\bibinfo  {journal} {JETP Lett.}\ }\textbf {\bibinfo
  {volume} {71}},\ \bibinfo {pages} {6} (\bibinfo {year} {2000})}\BibitemShut
  {NoStop}%
\bibitem [{\citenamefont {Lev}\ \emph {et~al.}(2001)\citenamefont {Lev},
  \citenamefont {Nazarenko}, \citenamefont {Nych}, \citenamefont {Schur},
  \citenamefont {Tomchuk}, \citenamefont {Yamamoto},\ and\ \citenamefont
  {Yokoyama}}]{Lev2001}%
  \BibitemOpen
  \bibfield  {author} {\bibinfo {author} {\bibfnamefont {B.}~\bibnamefont
  {Lev}}, \bibinfo {author} {\bibfnamefont {V.}~\bibnamefont {Nazarenko}},
  \bibinfo {author} {\bibfnamefont {A.}~\bibnamefont {Nych}}, \bibinfo {author}
  {\bibfnamefont {D.}~\bibnamefont {Schur}}, \bibinfo {author} {\bibfnamefont
  {P.}~\bibnamefont {Tomchuk}}, \bibinfo {author} {\bibfnamefont
  {J.}~\bibnamefont {Yamamoto}}, \ and\ \bibinfo {author} {\bibfnamefont
  {H.}~\bibnamefont {Yokoyama}},\ }\href@noop {} {\bibfield  {journal}
  {\bibinfo  {journal} {Phys. Rev. E}\ }\textbf {\bibinfo {volume} {64}},\
  \bibinfo {pages} {021706} (\bibinfo {year} {2001})}\BibitemShut {NoStop}%
\bibitem [{\citenamefont {Auernhammer}\ \emph {et~al.}(2009)\citenamefont
  {Auernhammer}, \citenamefont {Zhao}, \citenamefont {Ullrich},\ and\
  \citenamefont {Vollmer}}]{Auernhammer2009}%
  \BibitemOpen
  \bibfield  {author} {\bibinfo {author} {\bibfnamefont {G.~K.}\ \bibnamefont
  {Auernhammer}}, \bibinfo {author} {\bibfnamefont {J.}~\bibnamefont {Zhao}},
  \bibinfo {author} {\bibfnamefont {B.}~\bibnamefont {Ullrich}}, \ and\
  \bibinfo {author} {\bibfnamefont {D.}~\bibnamefont {Vollmer}},\ }\href@noop
  {} {\bibfield  {journal} {\bibinfo  {journal} {Eur. Phys. J. E}\ }\textbf
  {\bibinfo {volume} {30}},\ \bibinfo {pages} {4} (\bibinfo {year}
  {2009})}\BibitemShut {NoStop}%
\bibitem [{\citenamefont {Wagner}(1914)}]{Wagner1914}%
  \BibitemOpen
  \bibfield  {author} {\bibinfo {author} {\bibfnamefont {K.~W.}\ \bibnamefont
  {Wagner}},\ }\href@noop {} {\bibfield  {journal} {\bibinfo  {journal} {Archiv
  f{\"u}r Elektrotechnik}\ }\textbf {\bibinfo {volume} {2}},\ \bibinfo {pages}
  {371} (\bibinfo {year} {1914})}\BibitemShut {NoStop}%
\bibitem [{\citenamefont {Doostmohammadi}\ \emph {et~al.}(2016)\citenamefont
  {Doostmohammadi}, \citenamefont {Thampi},\ and\ \citenamefont
  {Yeomans}}]{DoostmohammadiPRL2016}%
  \BibitemOpen
  \bibfield  {author} {\bibinfo {author} {\bibfnamefont {A.}~\bibnamefont
  {Doostmohammadi}}, \bibinfo {author} {\bibfnamefont {S.~P.}\ \bibnamefont
  {Thampi}}, \ and\ \bibinfo {author} {\bibfnamefont {J.~M.}\ \bibnamefont
  {Yeomans}},\ }\href@noop {} {\bibfield  {journal} {\bibinfo  {journal} {Phys.
  Rev. Lett.}\ }\textbf {\bibinfo {volume} {117}},\ \bibinfo {pages} {048102}
  (\bibinfo {year} {2016})}\BibitemShut {NoStop}%
\bibitem [{\citenamefont {{De Gennes}}\ and\ \citenamefont
  {Prost}(1995)}]{DeGennes1995}%
  \BibitemOpen
  \bibfield  {author} {\bibinfo {author} {\bibfnamefont {P.}~\bibnamefont {{De
  Gennes}}}\ and\ \bibinfo {author} {\bibfnamefont {J.}~\bibnamefont {Prost}},\
  }\href@noop {} {\emph {\bibinfo {title} {{The Physics of Liquid
  Crystals}}}},\ \bibinfo {edition} {2nd}\ ed.\ (\bibinfo  {publisher} {Oxford
  University Press},\ \bibinfo {address} {Oxford},\ \bibinfo {year}
  {1995})\BibitemShut {NoStop}%
\bibitem [{\citenamefont {Orlandini}\ \emph {et~al.}(1995)\citenamefont
  {Orlandini}, \citenamefont {Swift},\ and\ \citenamefont
  {Yeomans}}]{Orlandini1995a}%
  \BibitemOpen
  \bibfield  {author} {\bibinfo {author} {\bibfnamefont {E.}~\bibnamefont
  {Orlandini}}, \bibinfo {author} {\bibfnamefont {M.~R.}\ \bibnamefont
  {Swift}}, \ and\ \bibinfo {author} {\bibfnamefont {J.~M.}\ \bibnamefont
  {Yeomans}},\ }\href@noop {} {\bibfield  {journal} {\bibinfo  {journal}
  {Europhys. Lett.}\ }\textbf {\bibinfo {volume} {32}},\ \bibinfo {pages} {463}
  (\bibinfo {year} {1995})}\BibitemShut {NoStop}%
\bibitem [{\citenamefont {Matsuyama}\ \emph {et~al.}(2002)\citenamefont
  {Matsuyama}, \citenamefont {Evans},\ and\ \citenamefont
  {Cates}}]{Matsuyama2002}%
  \BibitemOpen
  \bibfield  {author} {\bibinfo {author} {\bibfnamefont {A.}~\bibnamefont
  {Matsuyama}}, \bibinfo {author} {\bibfnamefont {R.~M.~L.}\ \bibnamefont
  {Evans}}, \ and\ \bibinfo {author} {\bibfnamefont {M.~E.}\ \bibnamefont
  {Cates}},\ }\href@noop {} {\bibfield  {journal} {\bibinfo  {journal} {Eur.
  Phys. J. E}\ }\textbf {\bibinfo {volume} {9}},\ \bibinfo {pages} {1}
  (\bibinfo {year} {2002})}\BibitemShut {NoStop}%
\bibitem [{\citenamefont {Sulaiman}\ \emph {et~al.}(2006)\citenamefont
  {Sulaiman}, \citenamefont {Marenduzzo},\ and\ \citenamefont
  {Yeomans}}]{Sulaiman2006}%
  \BibitemOpen
  \bibfield  {author} {\bibinfo {author} {\bibfnamefont {N.}~\bibnamefont
  {Sulaiman}}, \bibinfo {author} {\bibfnamefont {D.}~\bibnamefont
  {Marenduzzo}}, \ and\ \bibinfo {author} {\bibfnamefont {J.~M.}\ \bibnamefont
  {Yeomans}},\ }\href@noop {} {\bibfield  {journal} {\bibinfo  {journal} {Phys.
  Rev. E}\ }\textbf {\bibinfo {volume} {74}},\ \bibinfo {pages} {041708}
  (\bibinfo {year} {2006})}\BibitemShut {NoStop}%
\bibitem [{\citenamefont {Cahn}\ and\ \citenamefont
  {Hilliard}(1958)}]{Cahn1958}%
  \BibitemOpen
  \bibfield  {author} {\bibinfo {author} {\bibfnamefont {J.~W.}\ \bibnamefont
  {Cahn}}\ and\ \bibinfo {author} {\bibfnamefont {J.~E.}\ \bibnamefont
  {Hilliard}},\ }\href@noop {} {\bibfield  {journal} {\bibinfo  {journal} {J.
  Chem. Phys.}\ }\textbf {\bibinfo {volume} {28}},\ \bibinfo {pages} {258}
  (\bibinfo {year} {1958})}\BibitemShut {NoStop}%
\bibitem [{\citenamefont {Beris}\ and\ \citenamefont
  {Edwards}(1994)}]{Beris1994}%
  \BibitemOpen
  \bibfield  {author} {\bibinfo {author} {\bibfnamefont {A.}~\bibnamefont
  {Beris}}\ and\ \bibinfo {author} {\bibfnamefont {B.}~\bibnamefont
  {Edwards}},\ }\href@noop {} {\emph {\bibinfo {title} {{Thermodynamics of
  Flowing Systems}}}}\ (\bibinfo  {publisher} {Oxford University Press},\
  \bibinfo {address} {Oxford},\ \bibinfo {year} {1994})\BibitemShut {NoStop}%
\bibitem [{\citenamefont {Paineau}\ \emph {et~al.}(2012)\citenamefont
  {Paineau}, \citenamefont {Dozov}, \citenamefont {Bihannic}, \citenamefont
  {Baravian}, \citenamefont {Krapf}, \citenamefont {Philippe}, \citenamefont
  {Rouzi{\`e}re}, \citenamefont {Michot},\ and\ \citenamefont
  {Davidson}}]{Paineau2012}%
  \BibitemOpen
  \bibfield  {author} {\bibinfo {author} {\bibfnamefont {E.}~\bibnamefont
  {Paineau}}, \bibinfo {author} {\bibfnamefont {I.}~\bibnamefont {Dozov}},
  \bibinfo {author} {\bibfnamefont {I.}~\bibnamefont {Bihannic}}, \bibinfo
  {author} {\bibfnamefont {C.}~\bibnamefont {Baravian}}, \bibinfo {author}
  {\bibfnamefont {M.-E.~M.}\ \bibnamefont {Krapf}}, \bibinfo {author}
  {\bibfnamefont {A.-M.}\ \bibnamefont {Philippe}}, \bibinfo {author}
  {\bibfnamefont {S.}~\bibnamefont {Rouzi{\`e}re}}, \bibinfo {author}
  {\bibfnamefont {L.~J.}\ \bibnamefont {Michot}}, \ and\ \bibinfo {author}
  {\bibfnamefont {P.}~\bibnamefont {Davidson}},\ }\href@noop {} {\bibfield
  {journal} {\bibinfo  {journal} {ACS Appl. Mater. Interfaces}\ }\textbf
  {\bibinfo {volume} {4}},\ \bibinfo {pages} {8} (\bibinfo {year}
  {2012})}\BibitemShut {NoStop}%
\end{thebibliography}%

\end{document}